\documentclass[10pt,sigconf,letterpaper]{acmart}
\AtBeginDocument{%
  }

\usepackage{graphicx}
\usepackage{float}
\usepackage{subfigure}

\usepackage{siunitx}
\DeclareSIUnit{\gbps}{Gb/s}
\DeclareSIUnit{\gb}{Gb}

\usepackage[nolist,nohyperlinks]{acronym}
\acrodefplural{LSE}[LSEs]{Label Stack Entries}

\usepackage{caption}
\acmYear{2025}\copyrightyear{2025}
\setcopyright{rightsretained}
\acmConference[ANRW 25]{Applied Networking Research Workshop}{July 22, 2025}{Madrid, Spain}
\acmBooktitle{Applied Networking Research Workshop (ANRW 25), July 22, 2025, Madrid, Spain}
\acmDOI{10.1145/3744200.3744779}
\acmISBN{979-8-4007-2009-3/25/07}

\newcommand\fig[1]{Figure~\ref{fig:#1}}
\newcommand\sect[1]{Section~\ref{sec:#1}}
\newcommand\equa[1]{Equation~\ref{eq:#1}}

\newcommand{\figeps}[3][]{%
  \begin{figure}[htb!]
    \centering
    \includegraphics[width=#1]{figures/#2}%
    \caption{#3}
    \Description{#3}
    \label{fig:#2}
  \end{figure}
}

\begin{document}

\begin{acronym}
    \acro{SR}{segment routing}
    \acro{ECMP}{equal-cost multi-path}
    \acro{OAM}{operations, administration and maintenance}
    \acro{SFC}{service function chaining}
	\acro{ISD}{in-stack data}
	\acro{PSD}{post-stack data}
	\acro{MPLS}{Multi Protocol Label Switching}
	\acro{MNA}{MPLS Network Actions}
	\acro{AD}{ancillary data}
	\acro{NAS}{Network Action Sub-stack}
	\acro{LSP}{Label Switched Path}
	\acro{LSE}{Label Stack Entry}
 	\acro{LSR}{Label Switching Router}
	\acro{LER}{Label Edge Router}
    \acro{bSPL}{base Special Purpose Label}
    \acro{eSPL}{extended Special Purpose Label}
    \acro{TC}{traffic class}
    \acro{FEC}{forwarding equivalence class}
	\acro{HBH}{hop-by-hop}
	\acro{I2E}{ingress-to-egress}
    \acro{RLD}{readable label depth}
	\acro{NAL}{Network Action Length}
    \acro{FRR}{Fast Reroute}
    \acro{NSH}{network service header}
 	\acro{P4}{Programming Protocol-independent Packet Processors}
   	\acro{MAT}{match+action table}
\end{acronym}

\title{Stack Management for MPLS Network Actions: Integration of Nodes with Limited Hardware Capabilities}


\author{Fabian Ihle}
\email{{fabian.ihle, menth}@uni-tuebingen.de}
\orcid{0009-0005-3917-2402}

\author{Michael Menth}
\orcid{0000-0002-3216-1015}
\affiliation{%
  \institution{Chair~of~Communication~Networks, University~of~Tuebingen}
  \city{Tuebingen}
  \country{Germany}}


\begin{abstract}
The MPLS Network Actions (MNA) framework enhances MPLS forwarding with a generalized encoding for manifold extensions such as network slicing and in-situ OAM (IOAM).
Network actions in MNA are encoded in \acp{LSE} and are added to the MPLS stack.
Routers have a physical limit on the number of \acp{LSE} they can read, called the \acf{RLD}.
With MNA, routers must be able to process a minimum number of \acp{LSE} which requires a relatively large \acs{RLD}.
In this paper, we perform a hardware analysis of an MNA implementation and identify the reason for a large RLD requirement in the MNA protocol design.
Based on this, we present a mechanism that reduces the required \acs{RLD} for MNA nodes by restructuring the MPLS stack during forwarding.
We then introduce the novel stack management network action that enables the proposed mechanism as well as its integration in networks with MNA-incapable nodes.
The feasibility of the mechanism on programmable hardware is verified by providing a P4-based implementation.
Further, the effects on the required \acs{RLD}, ECMP, and packet overhead are discussed.
\end{abstract}



\received{15 April 2025}
\received[accepted]{05 June 2025}

\maketitle

\section{Introduction}
The \acs{MPLS} protocol is a well-established forwarding technique for wide-area networking.
In MPLS, packets are encapsulated with labels that provide forwarding information specific to the network domain.
Label switching routers leverage these labels to forward packets to intermediate nodes.
Labels may be stacked to enable source routing or tunneling.

Over the last few decades, more sophisticated features have been added to the MPLS protocol such as \ac{ECMP}, IOAM, and \ac{SFC} encapsulation.
With these extensions, not only forwarding information but also information about the processing of a packet is carried in the MPLS stack.
This information is encoded using a special reserved range for label values.
However, the number of available reserved labels is limited~\cite{ietf-mpls-mna-fwk-05}.
This led to the standardization efforts of the IETF MPLS working group (WG) which created the MPLS Network Actions (MNA) framework for encoding network actions and their data in the MPLS stack.
Network actions in MPLS are comparable to extension headers in IPv6.
MNA is currently undergoing standardization in IETF~\cite{ietf-mpls-mna-fwk-05,ietf-mpls-mna-hdr-04,ietf-mpls-mna-usecases-03,ietf-mpls-mna-requirements-07}.

Network actions must be applied during forwarding and therefore require hardware support to keep up with the fast data rates of MPLS.
Routers have a physical limit on the number of \acfp{LSE} they can read.
This value is referred to as \acf{RLD} in the MNA context and results from the available hardware resources, e.g., memory for parsing.
Network actions in the MNA framework add multiple \acp{LSE} to the stack increasing the required \acs{RLD} in routers.
Therefore, routers require additional resources to implement network action processing.
However, the available resources on such hardware are limited.

A P4-based implementation of the MNA framework has shown that the proposed MNA processing is implementable on programmable hardware~\cite{IhMe24}.
However, resources are wasted for parsing irrelevant information of the MPLS stack due to the structure of this stack in the proposed MNA framework.
While the P4-based MNA implementation has enough resources to compensate for this, hardware with fewer resources may not.
Therefore, the allocation of required resources for network action processing must be reduced to enable efficient line rate data plane processing across different hardware platforms.

The contribution of this paper is manifold.
First, we perform a hardware analysis of an MNA implementation to identify the problem of wasted resources arising from header stacking in the MNA protocol design.
Then, we propose the \acf{HBH} preservation mechanism which reduces the required \acs{RLD} for processing network actions by restructuring the MPLS stack during forwarding.
Next, we describe a problem and a solution with the \acs{HBH} preservation mechanism in networks with MNA-incapable nodes.
We then propose a novel MNA network action that instructs nodes to reorganize the MPLS stack after network action processing, called the stack management network action.
This network action enables the \ac{HBH} preservation mechanism as well as the integration of the mechanism in networks with MNA-incapable nodes.
We implement the HBH preservation mechanism and the network action in P4 on the Intel Tofino\texttrademark\ 2 switching ASIC.
Finally, we discuss the implications for the required RLD, \acs{ECMP}, and packet overhead caused by the network action.
\section{The MPLS Network Actions (MNA) Framework}
In this section, we describe the \ac{MNA} framework as proposed by the IETF MPLS working group~\cite{ietf-mpls-mna-fwk-05, ietf-mpls-mna-hdr-04}.
First, we give an overview, then we explain scopes in \ac{MNA}, and finally, we explain constraints from the \acf{RLD}.

\subsection{Overview}
The \ac{MNA} framework introduced by the IETF MPLS WG extends the MPLS protocol with a generalized encoding for transporting and processing network actions and their data~\cite{ietf-mpls-mna-fwk-05}.
Network actions are carried as \ac{ISD} or as \ac{PSD}~\cite{ietf-mpls-ps-mna-hdr-00}.
In this work, we only consider \acs{ISD} where network actions are inserted into the MPLS stack.
Those network actions are arranged in a so-called \acf{NAS}~\cite{ietf-mpls-mna-hdr-04}.
In a \acs{NAS}, the encoding of \acp{LSE} is repurposed to enable more sophisticated extensions.
A \acs{NAS} is at minimum two \acp{LSE} and at most 17 \acp{LSE} large, i.e., up to \qty{68}{\byte} (bytes).
There may be multiple \acs{NAS} present in an MPLS stack.
If a \acs{NAS} is exposed to the top, i.e., if the preceding forwarding label is popped, the \acs{NAS} must be popped.
A more detailed technological overview of the \acs{MNA} framework can be found in \cite{IhMe24}.

\subsection{Scopes in MNA}
\label{sec:scopes}

Network actions in MNA can be processed either on selected nodes or on all nodes on the path.
To that end, a \acs{NAS} has a scope indicated by a field in the \acs{NAS} encoding.
The scope of a \acs{NAS} defines on which nodes the network actions are processed.
Various \acs{NAS} with different scopes can be provided.
This concept is illustrated in \fig{pdfs/scopes} and further explained in the following\footnote{The MPLS forwarding examples in this paper comprise one forwarding label per node in an SR-MPLS fashion to clarify the challenges arising from the combination of SR-MPLS and MNA. In reality, however, a stack with only two forwarding labels for SR-MPLS in a network may suffice~\cite{BrSc25}.}.

\figeps[0.8\columnwidth]{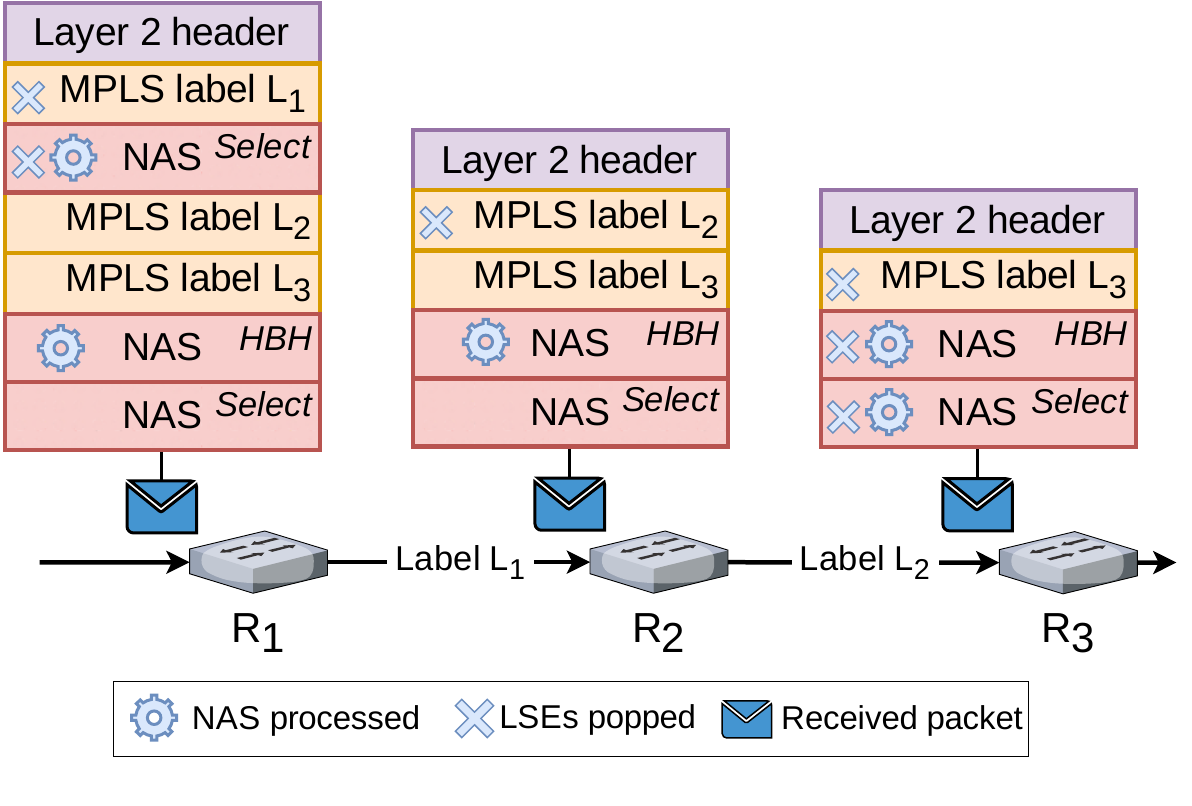}{The select-scoped and the HBH-scoped \acs{NAS} in MNA~\cite{IhMe24,ietf-mpls-mna-hdr-04}.}

The scopes include a select scope and a \acf{HBH} scope.
A \textit{select}-scoped NAS is only processed by one specific node. 
It is located below the forwarding label for that specific node and is only processed when the top-of-stack label is popped.
A select-scoped \acs{NAS} is always popped after processing~\cite{ietf-mpls-mna-hdr-04}.

An \textit{\acs{HBH}}-scoped \acs{NAS} must be processed by every node on the path and should not be removed by intermediate nodes.
By placing an \acs{HBH}-scoped \acs{NAS} at the bottom-of-stack, the \acs{NAS} is available to all nodes on the path because it is only exposed to the top and therefore popped by the last node.
However, this is not feasible for practical deployment.
This problem is further discussed in \sect{hbh_copies}.

In \fig{pdfs/scopes}, packets are forwarded according to the top-of-stack forwarding label.
Each node searches the MPLS stack for network actions it must process.
Node $R_1$ processes its select-scoped and the HBH-scoped \acs{NAS}, node $R_2$ processes the HBH-scoped NAS, and node $R_3$ processes the HBH-scoped and its select-scoped NAS.

\subsection{Constraints from the Readable Label Depth (RLD)}
\label{sec:hbh_copies}
The \acf{RLD} indicates the number of \acp{LSE} a node can read without performance impact and results from the available hardware resources.
Network actions must be within the \acs{RLD} of a node for processing.
Each node signals its \acs{RLD} to the ingress router using routing protocols such as IS-IS~\cite{rfc9088} or OSPF~\cite{rfc9089}.

In the previous example in \fig{pdfs/scopes}, it is assumed that the \acs{RLD} of each node is large enough to read the entire MPLS stack.
However, with limited hardware resources and a single NAS being up to 17 \acp{LSE} large, the entire MPLS stack is generally not within the \acs{RLD}.
An example with HBH-scoped NAS and a limited RLD based on ~\cite{IhMe24, ietf-mpls-mna-hdr-04} is illustrated in \fig{pdfs/hbh_copies}.

\figeps[0.8\columnwidth]{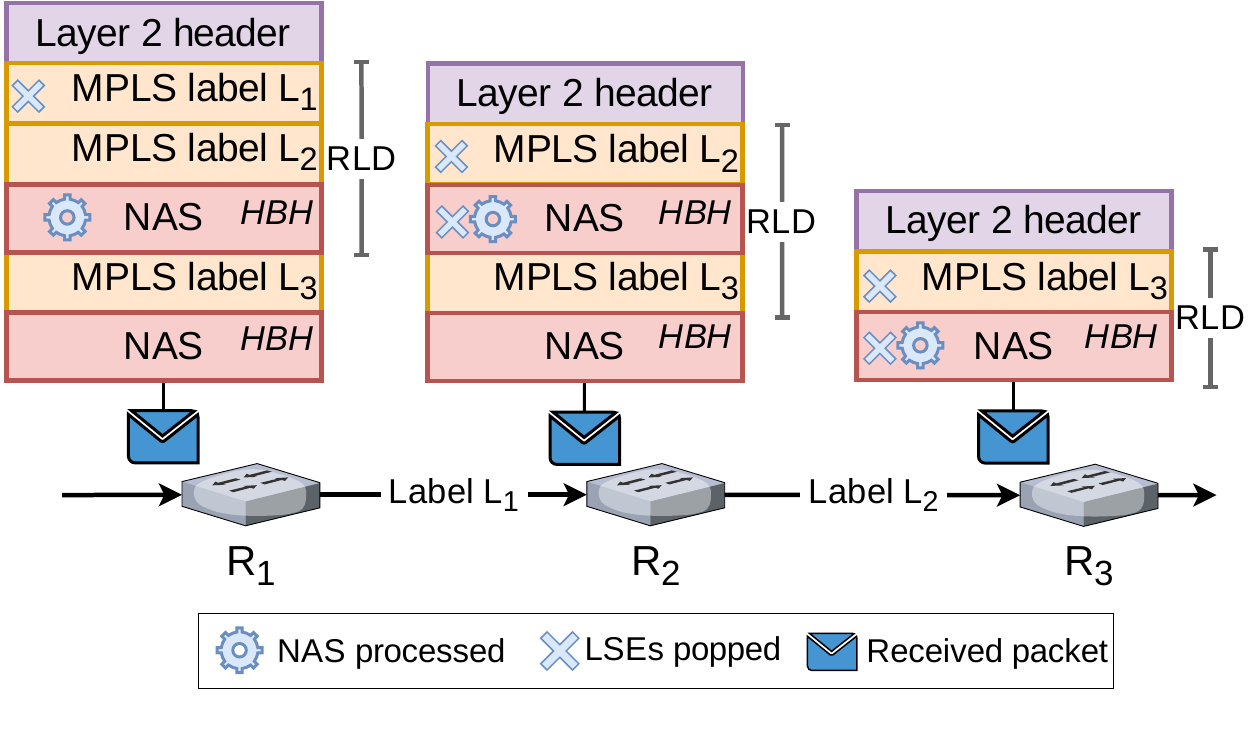}{A copy of the HBH-scoped NAS is placed in the MPLS stack to ensure that it is within RLD for each node on the path~\cite{IhMe24,ietf-mpls-mna-hdr-04}.}

In \fig{pdfs/hbh_copies}, the \acs{RLD} is limited to three \acp{LSE} and nodes are not capable of parsing the entire MPLS stack\footnote{The concept of the \acs{RLD} in this example is simplified because a \acs{NAS} is treated as one \acs{LSE} and only \acs{HBH}-scoped \acs{NAS} are shown.
In reality, a single \acs{NAS} is up to 17 \acp{LSE} large and the MPLS stack may contain multiple select-scoped \acs{NAS}.}.
An \acs{HBH}-scoped \acs{NAS} must be processed by every node on the path.
To that end, the NAS is placed at the bottom-of-stack in \fig{pdfs/scopes}.
However, with an \acs{RLD} of three \acp{LSE} and three forwarding labels in \fig{pdfs/hbh_copies}, a \acs{NAS} placed at the bottom-of-stack would be out of reach for node $R_1$.
The ingress router, i.e., the node pushing the MPLS stack including the \acs{NAS}, must ensure that the \acs{NAS} is within \acs{RLD} when the packet reaches the MNA-capable node~\cite{ietf-mpls-mna-hdr-04}.
Therefore, the ingress router has to place a copy of the HBH-scoped \acs{NAS} in the stack.
Only the first HBH-scoped NAS found is processed by a node.
The HBH NAS copy is popped after exposing it to the top.

\section{Related Work}
In this section, we describe related work about reducing the size of MPLS stacks.
The authors of~\cite{GiCa15, LaBr15, GuDu16, BrSc25, GuDu17} propose various algorithms for efficient path label encoding that effectively reduce the size of the label stack.
These algorithms optimize path computation to minimize the number of labels in the stack, thereby reducing the required RLD.
However, while these approaches successfully reduce the size of label stacks, they primarily optimize path computation and are therefore only applicable to forwarding labels.
Although reducing the number of forwarding labels can indirectly decrease the number of \acp{LSE} associated with network actions, it does not eliminate the fundamental challenge posed by MNA: the increase in required \acs{RLD} for routers processing network actions.
Our work addresses this particular challenge by introducing the HBH preservation mechanism which alleviates the increase in required \acs{RLD} for \acs{MNA}.

\section{Hardware Analysis of an MNA Implementation}
In this section, we analyze the array allocation for MNA in hardware based on the implementation by Ihle \textit{et al.}~\cite{IhMe24}.
Using this analysis, we define a problem statement that identifies the reason for a large \acs{RLD} requirement.

\subsection{Array Allocation}
To parse the relevant parts of the MPLS stack including network actions in hardware, separate arrays are allocated for parsing \acs{NAS}, i.e., one array per \acs{NAS} scope~\cite{IhMe24}.
These are illustrated in \fig{pdfs/isd_stacks}.

\figeps[0.55\columnwidth]{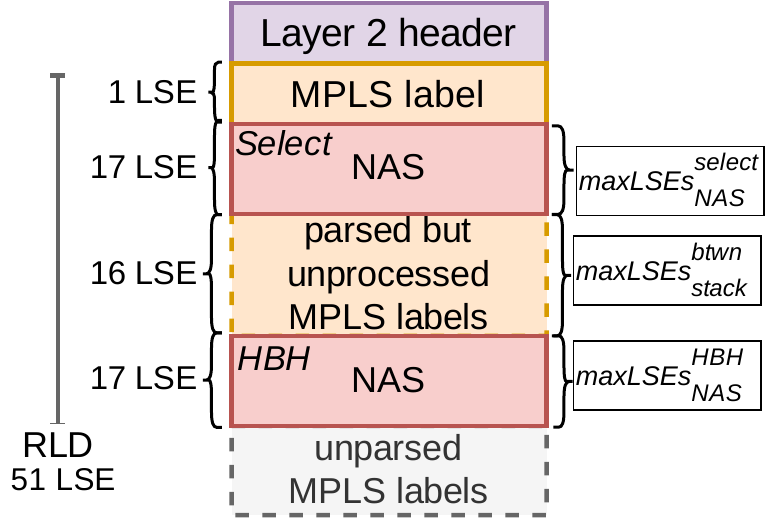}{Allocated arrays and their maximum sizes in the P4-based MNA implementation by Ihle \textit{et al.}~\cite{IhMe24}.}

There are three different arrays which hold a select-scoped \acs{NAS}, an \acs{HBH}-scoped \acs{NAS}, and arbitrary \acp{LSE} in between.
Nodes must parse \acp{LSE} of the \acs{MPLS} stack until they find the \acs{HBH}-scoped \acs{NAS} in the stack.
Those parsed \acp{LSE} are referred to as the in-between stack in~\cite{IhMe24}.
The arrays have to be allocated a priori and have a maximum size of $maxLSEs_{NAS}^{select}$, $maxLSEs_{NAS}^{HBH}$, and $maxLSEs_{stack}^{btwn}$.

According to the header encoding in the MNA framework, $maxLSEs_{NAS}^{select}$ and $maxLSEs_{NAS}^{HBH}$ are 17 \acp{LSE}~\cite{ietf-mpls-mna-hdr-04}.
Therefore, to support a select-scoped and an \acs{HBH}-scoped \acs{NAS} at a node, 34 \acp{LSE} must be accommodated in the two corresponding arrays.
The available \acs{RLD} ultimately constrains the size of the in-between stack as shown in \equa{btwn}~\cite{IhMe24}.
\begin{equation}
\begin{aligned}
    \text{maxLSEs}^{btwn}_{stack} = RLD 
    &- \text{maxLSEs}^{select}_{NAS}\\
    &\quad - \text{maxLSEs}^{HBH}_{NAS} - 1.\label{eq:btwn}
\end{aligned}
\end{equation}
For example, with an available \acs{RLD} of 51 \acp{LSE} in~\cite{IhMe24}, 17 \acp{LSE} remain to encode other \acp{LSE} above the HBH-scoped \acs{NAS}.
One of those \acp{LSE} is required for the top-of-stack forwarding label.
The remaining \acp{LSE} therefore allow for an in-between stack of $maxLSEs_{stack}^{btwn} = 16$ \acp{LSE}.
\acp{LSE} in the in-between stack encode forwarding labels or select-scoped \acs{NAS} for processing on other nodes.
With a smaller in-between stack, i.e., a smaller \acs{RLD}, there is not enough space in the header stack for other select-scoped \acs{NAS}.
The need for resource allocation of an in-between stack originates from the protocol design of the MNA framework and is not specific to the implementation in~\cite{IhMe24}.

\subsection{Problem Statement}
The problem leading to a large \acs{RLD} requirement originates from the processing of network actions and the structuring of \acs{NAS}.
Space for the maximum number of \acp{LSE} in the \acs{NAS} must be allocated.
Further, the \acs{HBH}-scoped \acs{NAS} is located somewhere deeper in the stack meaning that a node must parse a number of \acp{LSE} (the in-between stack) until it finds the \acs{HBH}-scoped \acs{NAS}.
The \acp{LSE} in the in-between stack are irrelevant for the current node and contain forwarding labels or network actions determined for other nodes.
However, they have to be parsed and resources have to be allocated in hardware for them.

Not having an in-between stack, i.e., placing an \acs{HBH}-scoped \acs{NAS} near the top of the stack, is not feasible in the current MNA proposal because that \acs{NAS} is popped before reaching all nodes and the \acs{HBH} \acs{NAS} must be preserved.
Therefore, in this work, we propose a mechanism that eliminates the need for the in-between stack by restructuring the MPLS stack during forwarding.
\section{The HBH Preservation Mechanism}
In this section, we first introduce the concept of the HBH preservation mechanism.
Next, we describe how backward compatibility with MNA-incapable nodes is ensured.
Finally, we propose a novel stack management network action to enable the HBH preservation mechanism and the integration of the mechanism in networks with MNA-incapable nodes.

\subsection{Concept of HBH Preservation}
\label{sec:concept}
As the \acp{LSE} contained in the in-between stack are not relevant to the current node, a better solution is to not have an in-between stack at all.
\fig{pdfs/mechanism} presents a mechanism to eliminate the need for an in-between stack while preserving the HBH NAS.
\figeps[0.7\columnwidth]{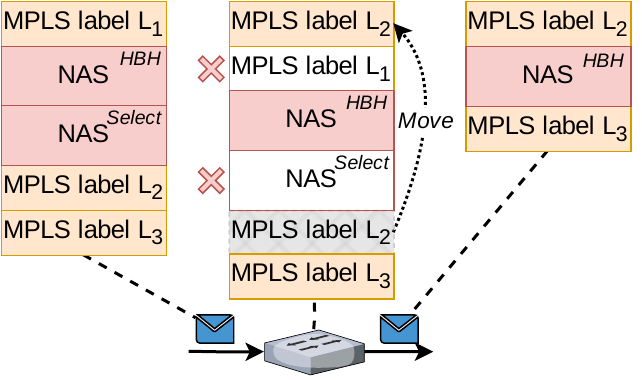}{
A node that pops the top-of-stack label moves the forwarding label below the \acs{NAS} to the top to preserve the \acs{HBH}-scoped \acs{NAS}.}
With the proposed HBH preservation mechanism, an \acs{HBH}-scoped \acs{NAS} is always located below the top-of-stack forwarding label.
However, after popping the forwarding label, the \acs{NAS} would be exposed to the top and therefore popped.
Thus, each MNA-capable node processing an HBH-scoped \acs{NAS} brings a forwarding label from below the \acs{NAS} to the top-of-stack if the top-of-stack forwarding label is popped.
This way, the \acs{HBH}-scoped \acs{NAS} is never exposed to the top, and the position of the \acs{HBH}-scoped \acs{NAS} in the stack is known, i.e., no in-between stack to locate the \acs{HBH}-scoped \acs{NAS} is required.

The HBH preservation mechanism is indicated in a network action  added to the HBH-scoped \acs{NAS}.
The network action is further described in \sect{network_action}.
With the network action, no state in transit nodes is required.
Further, transit nodes are not required to introduce new default forwarding behavior, and no changes are required in the MNA drafts.

\subsection{Ensuring Backward Compatibility}
\label{sec:backward}
First, we explain the problem with the HBH preservation mechanism in networks with MNA-incapable nodes and then, we propose a solution.
\subsubsection{Problem}
With the HBH preservation mechanism, a node brings the label below the \acs{NAS} to the top-of-stack for the next node to process.
However, if the next node is not capable of \acs{MNA}, it may pop the top-of-stack label and expose the \acs{HBH}-scoped \acs{NAS} to the top without removing or preserving it.
An example is illustrated in \fig{pdfs/legacy_problem}.
\figeps[0.75\columnwidth]{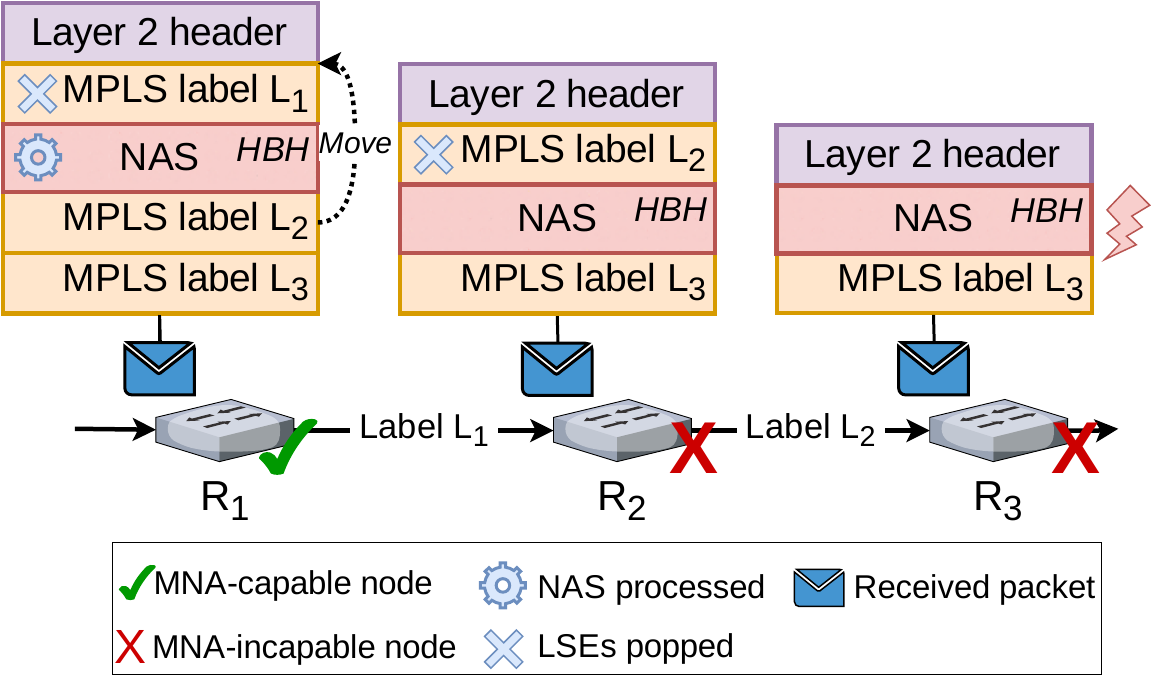}{A NAS is exposed to the top by MNA-incapable nodes leading to packet drop.}
As the \acs{MNA}-incapable node $R_2$ is not able to process the \acs{NAS}, the exposed \acs{NAS} is not popped.
The next MNA-incapable node $R_3$ will drop the packet because the top-of-stack label value, i.e., the MNA indicator, is invalid for an MNA-incapable node.
\subsubsection{Solution}
The ingress router has knowledge about the \acs{MNA} capabilities of each node on the path through signaling, e.g., via IS-IS or OSPF~\cite{ietf-mpls-mna-hdr-04}.
Therefore, the ingress router can instruct the preceding node of an MNA-incapable node to skip MNA processing on the succeeding MNA-incapable nodes.
This is achieved with the HBH preservation mechanism by bringing more than one forwarding label from below the NAS to the top-of-stack.
As the MNA capabilities of each node are signaled to the ingress router, no knowledge about the capabilities of neighboring nodes is required in transit nodes.
An example is given in \sect{network_action}.

\subsection{The Stack Management Network Action}
\label{sec:network_action}
We propose a network action that moves the next $n$ forwarding labels from below a \acs{NAS} to the top-of-stack.
We call this network action the stack management network action.
In the following, $n_{HBH}$ corresponds to the parameter $n$ indicated in an \acs{HBH}-scoped \acs{NAS} and $n_{Select}$ in a select-scoped \acs{NAS}, respectively.
The stack management network action brings $n = n_{HBH} + n_{Select}$ forwarding labels below the \acs{NAS} to the top if the top-of-stack forwarding label is popped.
If a forwarding label is not popped by a node, no labels need to be brought to the top.
The concept of the stack management network action enabling the HBH preservation mechanism with backward compatibility is illustrated in an example network in \fig{pdfs/example_backwards}.
\figeps[0.95\columnwidth]{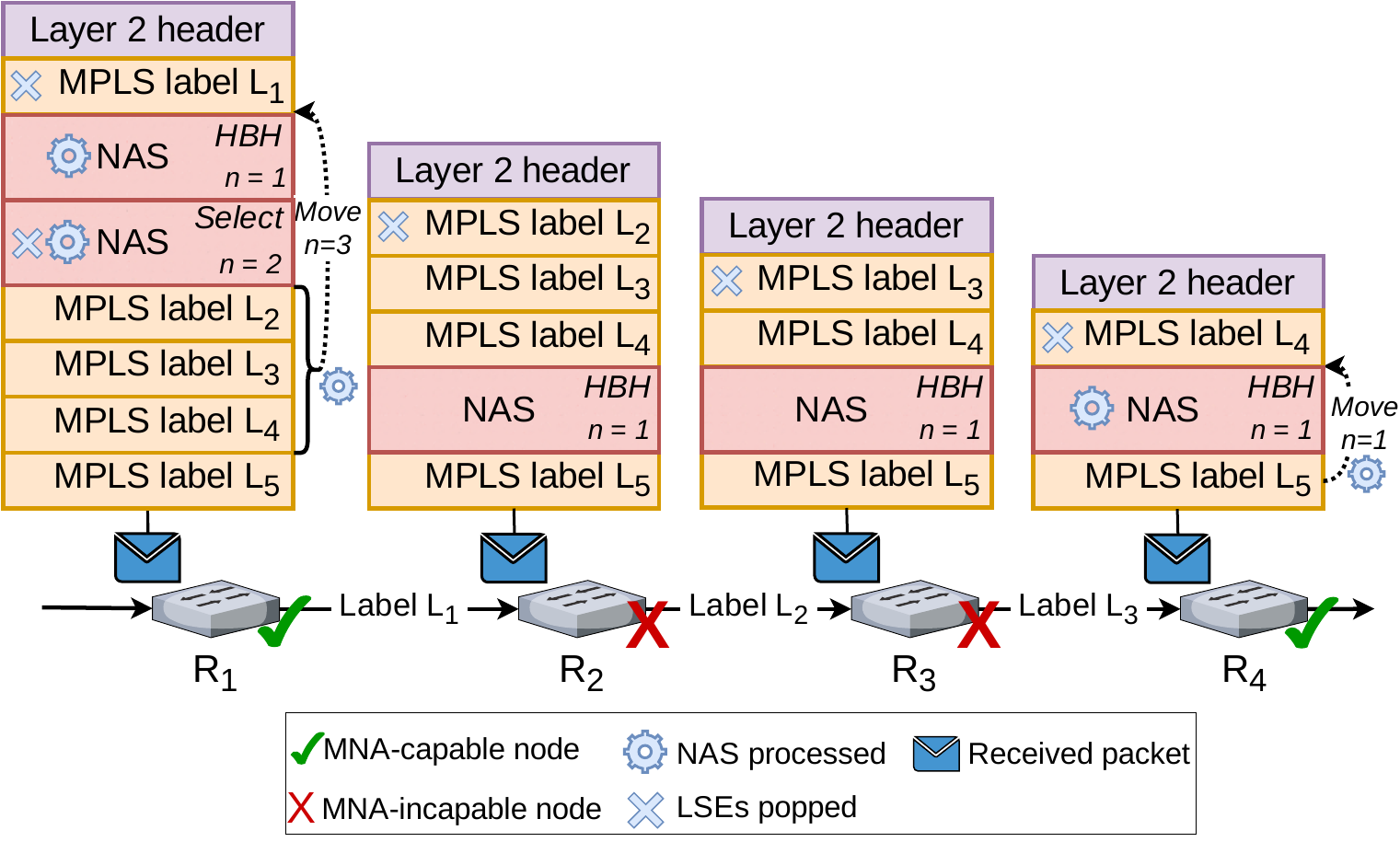}{An example network with two MNA-incapable nodes.}
In \fig{pdfs/example_backwards}, the stack management network action is added to the HBH-scoped \acs{NAS} with $n_{HBH}=1$ to perform the HBH preservation mechanism as described in \sect{concept}.
After node $R_1$, two MNA-incapable nodes follow.
For backward compatibility, the stack management network action is inserted in a select-scoped \acs{NAS} destined for the node preceding the MNA-incapable node, i.e., for $R_1$.
This allows $R_1$ to skip $n_{Select} = 2$ MNA-incapable nodes for backward compatibility as described in \sect{backward}.
In total, $R_1$ brings $n = n_{HBH} + n_{Select}=3$ labels to the top.
The select-scoped \acs{NAS} is popped after exposing it to the top at $R_1$, i.e., after popping the forwarding label $L_1$.
The MNA-incapable nodes $R_2$ and $R_3$ perform forwarding according to the top-of-stack label and do not expose the \acs{HBH}-scoped NAS to the top.
Finally, $R_4$ applies the HBH preservation mechanism with $n_{HBH}=1$.
A detailed Internet draft for the stack management network action is in preparation~\cite{draft-stack-management} and will be published to the IETF datatracker.

\section{Discussion}
In this section, we discuss the implementation feasibility of the proposed mechanism and describe the effect of the HBH preservation mechanism on the required \acs{RLD}.
Further, we discuss the header overhead and ECMP considerations.

\subsection{Proof of Concept Implementation}
We implemented the proposed stack management network action described in \sect{network_action} in P4 on the Intel Tofino\texttrademark\ 2 switching ASIC.
The P4 implementation of Ihle \textit{et al.} is extended for this purpose~\cite{IhMe24}.
The implementation of the HBH preservation mechanism including the proposed stack management network action operates at line rate of \SI{400}{\gbps} per port.
We verified the HBH preservation mechanism with a network as shown in \fig{pdfs/example_backwards} and conclude that an implementation of the proposed mechanism is feasible on programmable hardware.
The source code is available on GitHub~\cite{p4-mna-git}.
A detailed performance evaluation is left for future work.

A downside of the proposed approach is that the HBH preservation mechanism does not work on legacy MPLS hardware which is only capable of performing the traditional MPLS push, pop, and swap operations.
However, MNA processing on those nodes can be skipped using the proposed network action for backward compatibility.
Nodes capable of implementing network actions in the MNA framework, e.g., using P4-programmable hardware, can perform more sophisticated operations such as the stack management network action.

\subsection{Effect of HBH Preservation on the RLD}
With the \acs{HBH} preservation mechanism, a node is not required to search deeply in the stack for the \acs{HBH}-scoped \acs{NAS} as the \acs{NAS} is always located below the top-of-stack label.
The MPLS stack including the required minimum \acs{RLD} and the allocated array sizes at a node using the HBH preservation mechanism are shown in \fig{pdfs/new_rld}.
\figeps[0.7\columnwidth]{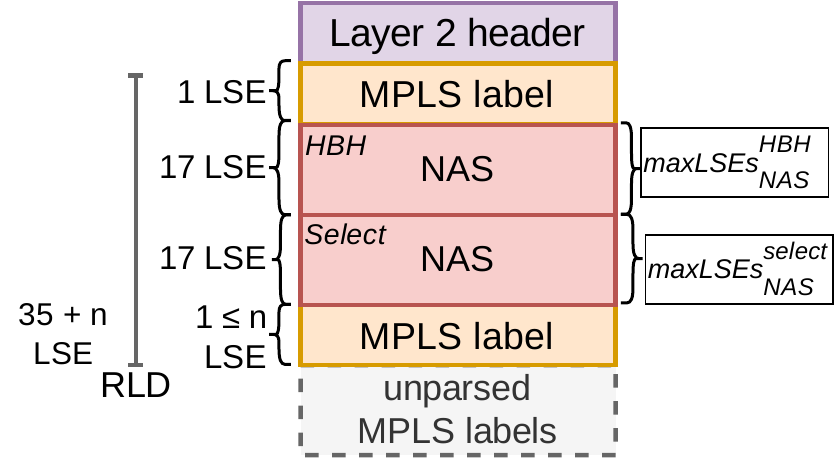}{Required minimum RLD with the HBH preservation mechanism.}
The required minimum RLD using the HBH preservation mechanism is $35 + n$ \acp{LSE} where $n \ge 1$.
Two maximum-sized \acs{NAS} with 17 \acp{LSE} each, one label for forwarding, and at least one label after the NAS are contained in the RLD.
The parameter $n$ corresponds to the number of \acp{LSE} that needs to be brought to the top at once, e.g., the maximum number of consecutive MNA-incapable nodes in a network. 
The proposed mechanism reduces the required allocation of resources by eliminating the need for an in-between stack.

Another advantage of the proposed mechanism is that the ingress router is not required to insert copies of the \acs{HBH}-scoped \acs{NAS} into the stack that are within the \acs{RLD} for each node as explained in \sect{hbh_copies}.
The \acs{HBH}-scoped \acs{NAS} is always close to the top-of-stack label and therefore within \acs{RLD} for each node.
This reduces the overall size of the stack.

The minimum \acs{RLD} of 36 \acp{LSE} in our approach may still be too large for hardware with fewer resources.
However, 35 \acp{LSE} is the absolute minimum for supporting maximum-sized \acs{NAS} at a node in the MNA framework\cite{ietf-mpls-mna-hdr-04, IhMe24}.
For hardware with fewer resources, the maximum supported size of a \acs{NAS} can be reduced and signaled as proposed in \cite{IhMe24} to further reduce the required \acs{RLD}.

Hardware with a larger \acs{RLD} than the required minimum of 36 \acp{LSE} is more flexible with the HBH preservation mechanism because it does not need to allocate resources for the in-between stack.
Instead, those resources can be used to facilitate a \acf{PSD} implementation.
For a \acs{PSD} implementation, more \acp{LSE} may need to be parsed, such as the post-stack itself and \acp{LSE} down to the bottom-of-stack.
A mixed implementation of \acs{ISD} and \acs{PSD} will be explored in the future.

\subsection{Packet Header Overhead}
The HBH preservation mechanism introduces additional packet header overhead because network actions are added to the \acs{NAS}.
When an \acs{HBH}-scoped \acs{NAS} is present, an additional \acs{LSE}, i.e., \qty{4}{\byte}, is added to each packet to encode the stack management network action.
If no \acs{HBH}-scoped \acs{NAS} is present, the stack management network action is not needed.
Additionally, if the stack management network action is applied for backward compatibility, one \acs{LSE} is added in a select-scoped \acs{NAS}.
The memory for the arrays is allocated for the entire \acs{NAS} regardless of present network actions in a packet.

\subsection{ECMP Considerations}
Reorganizing the MPLS stack can cause problems with ECMP load balancing.
Here, labels in the MPLS stack are hashed to determine the path through the network.
If labels are reorganized during forwarding, the hash value changes and therefore packets may take different paths.
However, if the HBH preservation mechanism is applied consistently to all packets, the hash value between packets and therefore, the path, does not change.
\section{Conclusion}
\label{sec:conclusion}
\acs{MNA} facilitates extensions by adding \acs{LSE} encodings to the MPLS stack.
The number of \acp{LSE} a router can read, i.e., the \acs{RLD}, is constrained by its physical resources.
Network actions must lie within a node’s \acs{RLD} for processing.
However, the current design of the MNA framework results in a large \acs{RLD} requirement.
In this paper, we performed a hardware analysis of an MNA implementation and identified the problem of wasted resources in an in-between stack resulting from the MNA protocol design.
Therefore, we proposed and implemented the HBH preservation mechanism.
The HBH preservation mechanism eliminates the need for an in-between stack by structuring and reorganizing the MPLS stack.
We introduced the stack management network action to apply the MPLS stack reorganization during forwarding.
This network action enables the HBH preservation mechanism as well as the integration of the mechanism in networks with MNA-incapable nodes.
We extended the P4-MNA implementation in~\cite{IhMe24} with the proposed network action.
The implementation applies the HBH preservation mechanism at a line rate of \SI{400}{\gbps} per port.
With the HBH preservation mechanism the required minimum \acs{RLD} is at least 36 \acp{LSE} where 35 \acp{LSE} are the absolute minimum to support maximum-sized \acs{NAS} in \acs{MNA}.
This mechanism reduces the required \acs{RLD} and the overall stack size, enabling the integration of hardware with fewer resources.
Further, hardware with a larger \acs{RLD} becomes more flexible and can use its resources for other purposes, such as a \acs{PSD} implementation.
We discussed the packet header overhead of the stack management network action which is negligible with \qty{4}{\byte} per action.
Further, we considered implications on ECMP and conclude that there is no impact on load balancing if the proposed mechanism is applied consistently.
\section*{Acknowledgment}
The authors acknowledge the funding by the Deutsche For-schungsgemeinschaft (DFG) under grant ME2727/3-1. 
The authors alone are responsible for the content of the paper.
\balance

\bibliographystyle{ACM-Reference-Format}
\bibliography{literature}

\end{document}